\documentclass[sigconf]{acmart}
\usepackage{balance}
\AtBeginDocument{%
  \providecommand\BibTeX{{%
    \normalfont B\kern-0.5em{\scshape i\kern-0.25em b}\kern-0.8em\TeX}}}

\setcopyright{acmcopyright}
\copyrightyear{2023} 
\acmYear{2023} 
\setcopyright{acmlicensed}\acmConference[MM '23]{Proceedings of the 31st ACM International Conference on Multimedia}{October 29-November 3, 2023}{Ottawa, ON, Canada}
\acmBooktitle{Proceedings of the 31st ACM International Conference on Multimedia (MM '23), October 29-November 3, 2023, Ottawa, ON, Canada}
\acmPrice{15.00}
\acmDOI{10.1145/3581783.3613434}
\acmISBN{979-8-4007-0108-5/23/10}

\begin{document}

\title{Encoding and Decoding Narratives: Datafication and Alternative Access Models for Audiovisual Archives}


\author{Yuchen Yang}
\affiliation{%
  \institution{Laboratory for Experimental Museology, EPFL}
  \streetaddress{Rte Cantonale}
  \city{Lausanne}
  \country{Switzerland}}


\begin{abstract}
  Situated in the intersection of audiovisual archives, computational methods, and immersive interactions, this work probes the increasingly important accessibility issues from a two-fold approach. Firstly, the work proposes an ontological data model to handle complex descriptors (metadata, feature vectors, etc.) with regard to user interactions. Secondly, this work examines text-to-video retrieval from an implementation perspective by proposing a classifier-enhanced workflow to deal with complex and hybrid queries and a training data augmentation workflow to improve performance. This work serves as the foundation for experimenting with novel public-facing access models to large audiovisual archives.
\end{abstract}



\maketitle

\section{Introduction}
This work focuses on real-world archives - Télévision Suisse Romande (RTS) - and aims to tackle the preservation and accessibility issue in the age of data. The work is divided into two intertwined sub-parts:

\textbf{A model for datafication.}
This part emphasises the preservation end for archives from a data perspective. The first goal is to propose an ontology to formally represent various levels and aspects of data from AV archives and interactive experiences. The proposed ontological model reflects the data lineage among technical, content, conceptual, and interaction (meta)data and nurtures an update of preservation practices from a data management perspective. Using the model as a starting point, this part of the research will propose a data system and schema addressing retrieval efficiency and effectiveness. Together, deliverables from this section will provide theoretical and practical support for the preservation and new explorative methods. 

\textbf{Encoding AV archive: text-to-video embedding.}
This part of the research focuses on solving accessibility issues by verifying and improving the state-of-the-art text-to-video methods for real-world archives. The objective for this part is to explore the usage of such novel models beyond standard datasets and search for specific videos, Fusing it into various workflows creates applicable solutions to support real-world applications that focus on meaningful explorations.

\section{Related work}
\subsection{Archives and datafication}
In recent years, AV archives started to experiment with a deeper operationalisation on the content level, building manual or automated tools for annotations \cite{arnold2021introduction}, and finding new ways of using content descriptors and the ever-growing surrounding knowledge. Some focus on themed analysis - using colour to analyse aesthetics \cite{flueckiger2020methods} or movements for choreography \cite{broadwell2021comparative}. More general ones focus on tool sets for semantic annotations for audiovisual content \cite{cooper2021exploring, williams2021media}.

With more diverse data, curatorial practices with AV archives shift to immersive and interactive experiences to explore the plurality of memory materials and encourage personalised sense-making \cite{kenderdine2021computational}. The Pods in the Eye Filmmuseum \cite{ingravalle2015remixing} aims to explore the remixing of historical archives and the SEMIA project \cite{masson2020exploring} works on experimenting with alternative archive interfaces. Commercial tools like the Storyformer \cite{ursu2020authoring} also become available for creating personalised and interactive content experiences.

However, such fragmented practices remain impulsive and unsystematic. Few seek to reconcile the content, technology, and curatorial needs into reusable solutions and there is a lacking of fundamental models for mapping, linking, and managing the growing complex data. 

\subsection{text-to-video embedding}
The recently popularised text-to-video retrieval task takes an arbitrary text query and searches for the most relevant video clips accordingly. First proposed in 2016 \cite{rohrbach2017movie}, the text-to-video embedding usually has a two-stream architecture \cite{miech2019howto100m, miech2020end} utilising videos and corresponding descriptions. Videos and texts are transformed by encoders into vectors and projected into a common feature space. The paired relationships of the vectors are then used for training (or transforming) this common feature space so that the paired vectors are the closest to each other. This trained common feature space, called joint embedding space, is then used for inference and retrieval. Bi-directional loss \cite{karpathy2014deep}, symmetric cross entropy loss\cite{wang2019symmetric}, and triplet loss \cite{schroff2015facenet} are often used to train a joint embedding space. The training of text-to-video retrieval models benefits from manually labelled datasets for language-to-video related tasks \cite{rohrbach2017movie, anne2017localizing, xu2016msr}. 

Some works focus on improving the encoding of video using multimodal cues such as the face, audio, and speech \cite{shvetsova2022everything, gabeur2020multi}, while others work on generalising such models with large-scale pretraining. \cite{miech2020end} uses machine-generated transcription as descriptions for videos to construct large training sets. \cite{dzabraev2021mdmmt, portillo2021straightforward, xue2022clip, kunitsyn2022mdmmt} rely on large language to image models to pretrain, and finetune the model with text-video datasets.

While methods are maturing, the ability to work with real-world archives and queries is unknown. Models' performance drops on datasets with sophisticated video descriptions \cite{ji2022cret}. A recent study has proven that improved annotations would improve models' performance \cite{chen2022msr}.

\section{Methods and experiments}
\subsection{A model for datafication} 
\begin{figure}[H]
    \centering
    \includegraphics[width=0.5\textwidth]{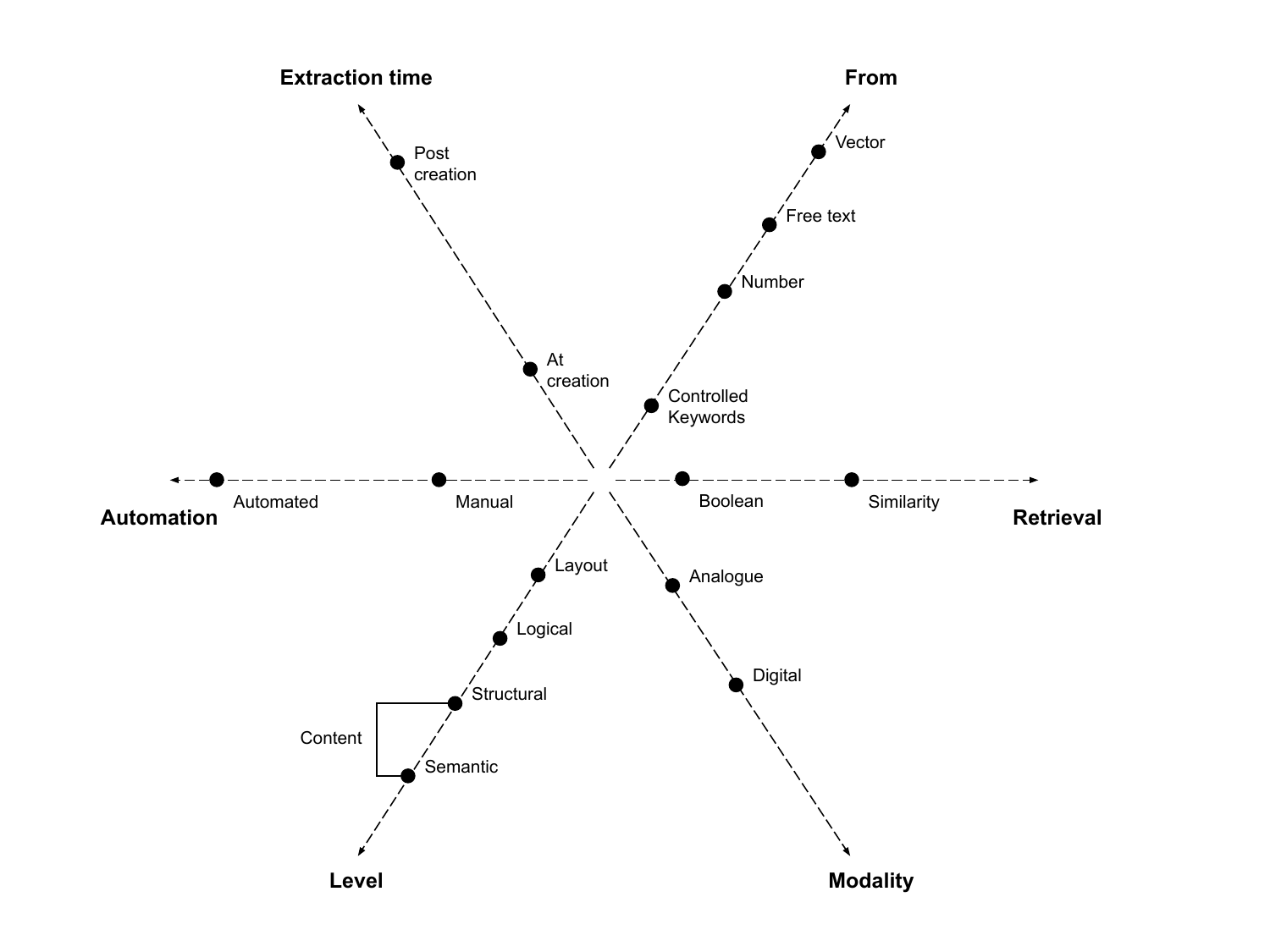}
    \caption[]{Dimensions of the metadata of a multimedia object visualised in the level, the automation, the extraction time, the form, the retrieval, and the modality dimensions}
    \label{fig:dim}
\end{figure}

Based on existing works \cite{de1998database, manjunath2002introduction, rossetto2021videograph}, a mapping
of the complexity of AV archive metadata in various dimensions is made and summarised in Figure \ref{fig:dim}.

Schemas like MPEG-7 \cite{manjunath2002introduction} are often used for such complexity in multimedia data and metadata management. However, they are not entirely suitable for content-level descriptors nowadays. Such schemas believe low-level metadata can be extracted automatically (like colour, texture, shape, timbre, pitch and rhythm), and high-level metadata are human annotations on a conceptual level (like emotions, content summaries) \cite{schallauer2011multimedia}. However, With maturing machine learning methods, the semantic gap of the inability to create high-level conceptual descriptors in an automated way is closing down \cite{wan2014deep}. Current standards have not reflected such changes. For example, descriptors such as body key points, text-to-video embedding vectors, and metadata (of training data, model, etc.) for machine learning methods are not considered at all. These are essential for understanding, improving, and managing the results, as well as a successful modern data management system \cite{kumar2017data, fourati2020survey}. 

On the other hand, with archival practices increasingly adapting interactive and personalised elements, only considering the content end is not enough. The public value of archives is realised through the combination of the \textbf{Content}, \textbf{Participants}, and \textbf{Interactions} \cite{carter2018narrative, austin2020narrative, dogan2020bringing}. However, the connection and relationship between these three parts have never been formalised. Inspired by previous works to formalise textual narratives using ontology \cite{bartalesi2016steps} and formalise the composition and relationship of multimedia big data \cite{rinaldi2018semantic}, this part of the research aims to provide an ontological model to describe the interconnection between the three aspects.

\subsection{Encoding AV archive: text-to-video embedding}
\subsubsection{Base text-to-video embedding model}
Since the purpose is to evaluate and improve the performance of text-to-video models when dealing with queries beyond plain visual description (such as with speech info). Models that do not consider audio information are discarded. Models are selected following these principles: representativeness, source code availability, and performance ranking. In the end, this work chooses the classic multimodal model MTT \cite{gabeur2020multi} and the latest multimodal retrieval model MFT \cite{shvetsova2022everything} as the core to construct the two proposed workflows.

\subsubsection{Base dataset}
This work uses the standard dataset MSR-VTT \cite{xu2016msr} as a base, which contains videos in music, sports, news, movie, drama, etc. The vastly diversified content mirrors the RTS archives the most. This dataset provides 10,000 video clips harvested from random internet sites, totalling 41.2 hours. Each video clip within this dataset is paired with 20 human annotations, contributing to 200,000 clip-sentence pairs.

\begin{figure}[h]
    \centering 
    \includegraphics[width=.8\columnwidth]{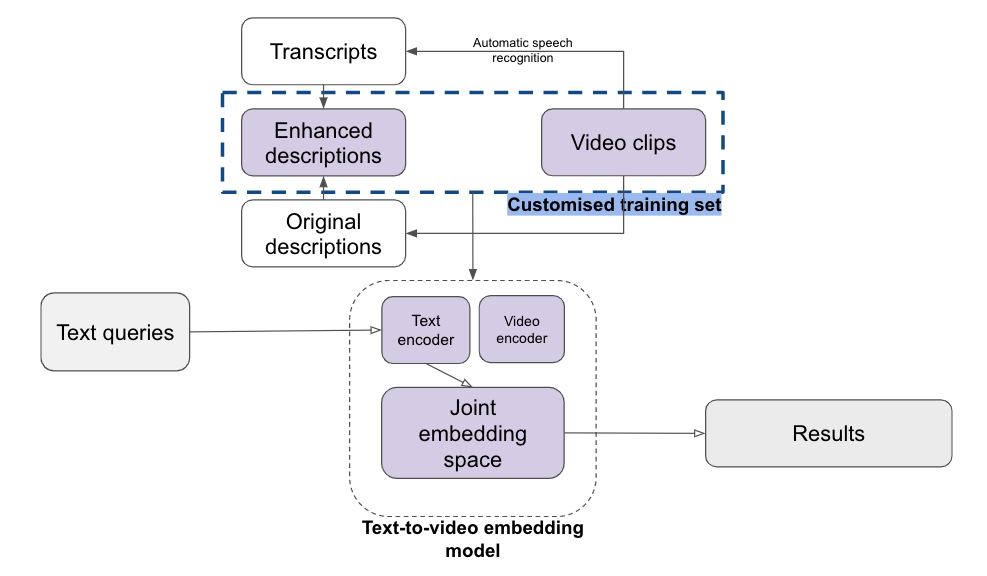} 
    \caption{A detailed look at the proposed single model with enhanced description workflow, the key component - customised training set - of this workflow is highlighted in the blue dashed box.}
\label{7}
\end{figure}

\subsubsection{Single model with enhanced description workflow} 
\textbf{Customised training set. }
Fig.\ref{7} depicts the workflow from end to end. The original annotations in the MSR-VTT are limited to plain and visual descriptions. For a video of a conversation between a contestant and judges, the annotations are "a girl and the judges talking on the voice" and "a girl is talking to the judges on a game show", which do not reflect the content of the conversation at all. To better understand if improving annotations would help with model performance, this workflow focuses on building a customised training set with the enhanced descriptions following a previous work using automatic speech recognition (ASR) \cite{miech2019howto100m}. The customised training set is contracted by randomly replacing one or more of the 20 original descriptions paired with a video clip with ASR results.

\textbf{Workflow implementation details. }
\textbf{Transcripts. }The MSR-VTT dataset is fed to Whisper \cite{radford2022robust} to obtain transcripts, using the provided "small model" and following the huggingface guides\footnote{\url{https://huggingface.co/openai/whisper-small}}. \textbf{Customised training set. }We use 1k-A split on MSR-VTT produced by \cite{yu2018joint} for constructing the customised training set. \textbf{Model Training. }The customised training set is then used to train the two base models, MMT and MFT, following their official implementation configurations respectively\footnote{MFT:\url{https://github.com/ninatu/everything_at_once}; MMT: \url{https://github.com/gabeur/mmt}}. This workflow utilises a customised training set and produces two new methods: customised MMT and customised MFT.

\begin{figure}[h]
    \centering 
    \includegraphics[width=.8\columnwidth]{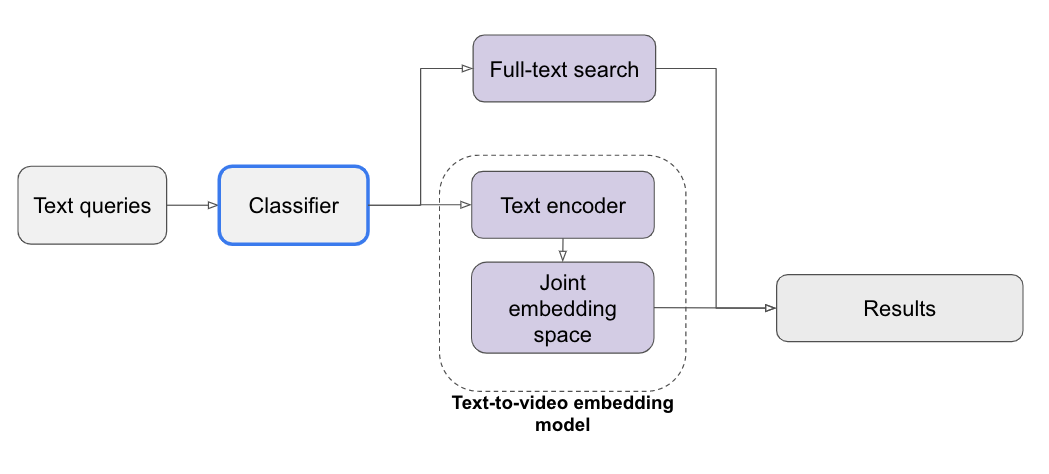} 
    \caption{A detailed look at the proposed classifier-enhanced workflow, the key component - classifier - of this workflow is highlighted in the blue box.}
\label{8}
\end{figure}

\subsubsection{Classifier-enhanced workflow}
\textbf{Classifier. }
As seen in Fig.\ref{8}, the fundamental idea is to classify queries and send them to the appropriate retrieval component. Although hard-coded rule sets can function to a certain extent, a machine-learning-based classifier is introduced in the hope of scalability and generalisation. In this work, the classifier distinguishes quote or speech-related texts from plain visual descriptions. Recent advancements in sequence models have brought significant transformations in natural language processing (NLP). Recurrent neural networks (RNNs) and transformers have consistently performed remarkably on numerous standard NLP benchmarks. We employ a long short-term memory (LSTM) architecture in this workflow to address the classification task.

\textbf{Workflow implementation details. }
\textbf{Training data for the classifier. }A labelled dataset with speeches or quotes and plain visual descriptions is prepared for the training. The labelled data for speeches or quotes are constructed by joining 2000 sentences from online databases\footnote{\url{https://libguides.bgsu.edu/c.php?g=227160&p=1505718}} using the regular expression filters converted advisors \footnote{\url{https://www.ccis.edu/student-life/advising-tutoring/writing-math-tutoring/introduce-quotations}}, and 1000 transcripts from the customised training set in Section 3.3.1. The labelled plain visual descriptions are 2000 random descriptions from the MSR-VTT 1k-A training set. An 80-20 split is done for training and testing. \textbf{Training the classifier. }The binary classifier is trained following previous work\footnote{\url{https://medium.com/holler-developers/intent-detection-using-sequence-models-ddae9cd861ee}} with the TensorFlow Keras sequential model. It contains an embedding layer representing each word with a vector length of 16. The following 16-unit LSTM layer uses relu activation. The final dense layer has seven units and a softmax activation for classification. The model is fit on the training dataset with a batch size of 32 for seven epochs. \textbf{Text-to-video model. } This workflow sends non-speech or non-quote queries to the text-to-video model. The original MSR-VTT 1k-A training set is used to train the two base text-to-video embedding models MMT and MFT, following their official implementation configurations respectively\footnote{MFT:\url{https://github.com/ninatu/everything_at_once}; MMT: \url{https://github.com/gabeur/mmt}}. \textbf{Full-text search model. }For quote- or speech-related queries, the workflow sends them to the full-text search. The ASR results from MSR-VTT in Section 3.3.1 are stored in an ElasticSearch\footnote{\url{https://www.elastic.co/}} database. The similarity is calculated by built-in API for full-text queries. This workflow utilises a classifier and produces two new methods based on the two base text-to-video embedding models: classifier MMT and classifier MFT.

\begin{figure}[H]
    \centering
    \includegraphics[width=0.5\textwidth]{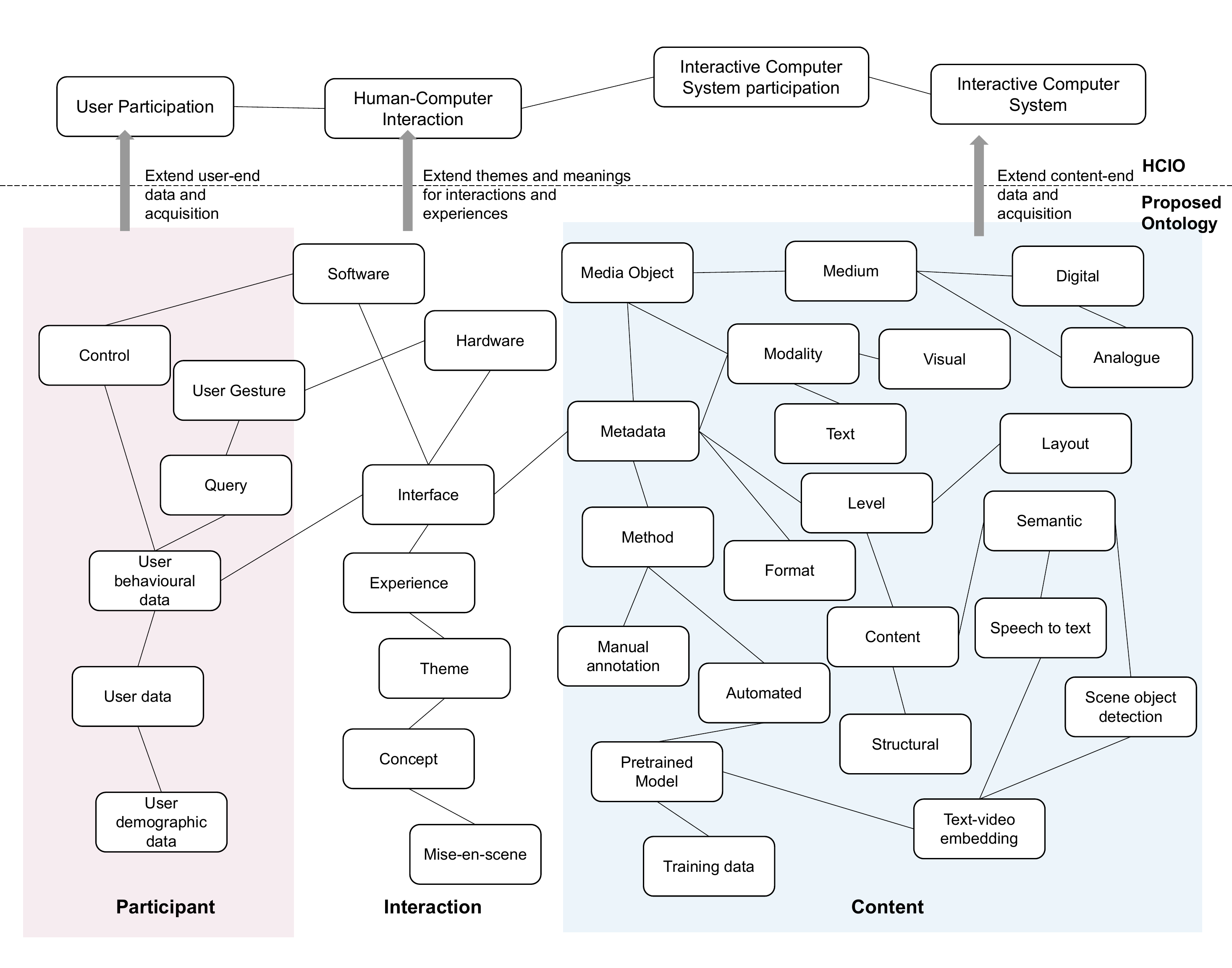}
    \caption[]{An exemplary look at some of its key classes and properties of it to illustrate the overall ontology concept}
    \label{fig:onto}
\end{figure}

\section{Preliminary RESULTS}
\subsection{A model for datafication}

This ontology is modulated along three corresponding dimensions - content, participant, and interaction (Fig. \ref{fig:onto})
The proposed ontology uses HCIO (\cite{costa2022core}) as a core and top-level reference for inheriting fundamental notions for describing the core aspects of the human-computer interaction phenomenon. Ontology for Media Resources\footnote{https://www.w3.org/TR/mediaont-10}, COMM (\cite{arndt2009comm}), and an ontology for harvesting
user input in the immersive environment (\cite{chera2012gesture}) are used as starting point to populate the content, participant, and interaction dimensions. 

\subsection{Encoding AV archive: text-to-video embedding}
\subsubsection{Test setup}
\textbf{Baseline test set. }The popular 1k-A split on MSR-VTT provides the baseline test dataset. \textbf{Customised test set. }A customised test set is introduced to evaluate different methods' performance in complex query situations better. The customised set is constructed on the base of the 1k-A split test set on MSR-VTT. We randomly replaced 50\% of the original ground truth (one random entry from the 20 annotations) with the obtained transcripts for those video clips. The result is 1,000 ground truth pairs with a mixture of plain visual descriptions and speeches or quotes.  \textbf{Evaluation metrics. }Standard matrics R@5 are likely to be more useful when understanding the performance and hence picked to report in the result.

\subsubsection{Comparison with state-of-the-art} 
Table \ref{table:modelresult} reports the result of all methods constructed for the evaluation on the comparison with the state-of-the-art. The baseline models' performance on the original MSR-VTT 1k-A train-test split is referenced when available from the original paper. Following the official implementations, the two models are also trained according to the original training settings from scratch using the 1k-A split of the original and the customised MSR-VTT dataset. The test sets from the 1k-A split of the original and the customised MSR-VTT dataset are used to conduct the final evaluation. Overall, the proposed Classifier MFT method achieved comparable performance with the baseline state-of-the-art model, with an R@5 of 54.2 compared to 57.1. On the customised MSR-VTT test set, where the query situation is a bit more complex, Classifier MFT outperforms all other methods with an R@5 of 77.5.

\begin{table}[h]
\scalebox{0.7}{
\begin{tabular}{cccc}
\hline
Method & Training Dataset & Original MSR-VTT & Customised MSR-VTT \\
                 &                    & R@5↑          & R@5↑          \\ \hline
MMT              & Original MSR-VTT   & 54.0          & 12.9          \\
MFT              & Original MSR-VTT   & \textbf{57.1} & 11.2          \\ \hline
Customised MMT   & Customised MSR-VTT & 49.5          & 19.0          \\
Customised MFT    & Customised MSR-VTT & 47.0          & \textbf{22.1} \\ \hline
Classifier MMT   & Original MSR-VTT   & 52.7          & 76.2          \\
Classifier MFT   & Original MSR-VTT                  & \textbf{54.2}    & \textbf{77.5}      \\ \hline
\end{tabular}}
\caption{Results of the baseline, customised, and classifier-enhanced method on the original and customised MSR-VTT test sets.}
\label{table:modelresult}
\end{table}

Several observations can be made based on the experiment results. First, all four baseline and customised methods suffer a performance drop when dealing with quote-related queries specific to speech information. This expected behaviour could be caused by the fact that most of the speech information is not matched with the visual perspective of the video clips in the given dataset. Second, models trained on the customised dataset, which contains descriptions that are transcriptions of the video, perform slightly worse when tested with the original test set, but better when dealing with a hybrid of descriptive and quote-related queries. Adding the speech-related descriptions in the customised dataset can provide more information during the training and slightly improve the performance when dealing with queries targeted more on the audio perspective. However, the extra information can also be regarded as noise, messing up the joint-embedding space and undermining the overall performance. Third, all classifier-enhanced methods perform well in both test sets. However, it is noticeable that the performance when dealing with the original test set drops slightly compared to the baseline methods. This can be caused by the fact that the performance is heavily determined by the classifier's performance, in which case it will not be 100\% accurate.

\section{Future work}
\textbf{A model for datafication}
The development of this ontology is an ongoing and circular task. Taking advantage of the Sinergia project, this ontology will be developed, validated, and improved with archival partners and exhibition installation. Another natural next step is to build a flexible data schema (JSON) to store the array of feature vectors and metadata with temporospatial consideration to facilitate better retrieval or explorative applications.

\textbf{Encoding AV archive: text-to-video embedding}
Multiple works on text-to-video retrieval methods have emphasised the importance of having a more situated and better quality dataset in improving the retrieval performance \cite{chen2022msr, shvetsova2022everything, fang2021clip2video}. In this specific work, only one additional quote-related query text is considered. However, narrative text, even as simple as a diary, has a much more diverse type of sentence describing many different aspects and levels of semantics within AV content. It would be beneficial to dig deeper in that direction and find a better strategy to create an appropriate customised annotation for video clips to reflect that. For instance, Vlogs, with the speech information being diverse enough to include many aspects of the given video, could be a more suitable source of descriptions for creating the text-video pairs to cover a more diverse scenario in the query \cite{fouhey2018lifestyle}. However, if the more complex annotation will be regarded as noise and hinder the performance is yet to be tested. New training strategies and architectures to handle the weakly-paired text are also required.

Together, the two parts of this research serve as the foundation and necessary support for exploring new access models for large AV archives from a novel and data-driven perspective. On top of this basis, the thesis will move on to build and evaluate prototypes for innovative experiences featuring more personalized, intuitive, serendipitous, and human-centric explorations of AV archives.
\begin{acks}
This research is made possible through the SNSF Sinergia project, grant number CRSII5\_198632.
\end{acks}
\newpage
\bibliographystyle{ACM-Reference-Format}
\balance 
\bibliography{sample-authordraft}

\end{document}